\documentclass[prl, 12pt, showpacs, showkeys]{revtex4}
\usepackage{amsfonts}
\usepackage{amscd}
\usepackage{amssymb}
\usepackage{amsmath}
\usepackage{latexsym}
\usepackage{mathrsfs}

\renewcommand{\tilde}{\widetilde}
\newcommand{\be}{\begin{equation}}
\newcommand{\ee}{\end{equation}}

\newcommand{\ba}{\begin{eqnarray}}
\newcommand{\ea}{\end{eqnarray}}

\newcommand{\del}{\partial}
\newcommand{\delbar}{{\overline{\del}}}

\newcommand{\bC}{{\mathbb{C}}}
\newcommand{\bE}{{\mathbb{E}}}

\newcommand{\bI}{{\mathbb{I}}}

\newcommand{\bP}{{\mathbb{P}}}

\newcommand{\bT}{{\mathbb{PT}}}

\newcommand{\cA}{{\mathcal{A}}}
\newcommand{\cB}{{\mathcal{B}}}

\newcommand{\cF}{{\mathcal{F}}}

\newcommand{\cO}{{\mathcal{O}}}

\newcommand{\rd}{{\rm d}}
\newcommand{\rD}{{\rm D}}

\newcommand{\e}{{\rm e}}
\newcommand{\im}{{\rm i}}

\begin{document}
\title{From Twistor Actions to MHV Diagrams}
\author{Rutger Boels, Lionel Mason \& David Skinner}
\affiliation{The Mathematical Institute, University of Oxford\\
24-29 St.~Giles, Oxford OX1 3LP, United Kingdom}
\begin{abstract}
  We show that MHV diagrams are the Feynman diagrams of certain
  twistor actions for gauge theories in an axial gauge. The gauge
  symmetry of the twistor action is larger than that on space-time and
  this allows us to fix a gauge that makes the MHV formalism manifest
  but which is inaccessible from space-time. The framework is
  extended to describe matter fields: as an illustration we explicitly
  construct twistor actions for an adjoint scalar with arbitrary
  polynomial potential and a fermion in the fundamental representation
  and show how this leads to additional towers of MHV vertices in the
  MHV diagram formalism.
\end{abstract}


\maketitle

An important outcome of Witten's twistor-string
theory~\cite{Witten:2003nn} is the ``MHV
formalism"~\cite{Cachazo:2004kj}, in which scattering amplitudes in
four dimensional Yang-Mills theory are described in terms of diagrams
whose vertices are the MHV vertices with two positive and arbitrarily
many negative helicity gluons.
Much work (see {\it
e.g.}~\cite{Bedford:2004nh,Georgiou:2004by,Birthwright:2005ak,Berger:2006vq})
has since gone into developing unitarity methods to compute loop
amplitudes, or on extending the MHV formalism to include matter
coupled to the Yang-Mills field~\cite{Dixon:2004za,Badger:2005zh}.

A key question here is the connection between these twistor-inspired
developments and the usual, Lagrangian-based approach to gauge theory.
Lagrangians have the obvious advantages of leading to a systematic
perturbation theory including loops, and of making manifest the symmetry properties of a theory. Progress towards a full derivation of the MHV formalism from the standard space-time Lagrangian in lightcone gauge has been made in~\cite{Gorsky:2005sf,Mansfield:2005yd,
  Ettle:2006bw, Feng:2006yy, Brandhuber:2006bf}. The key idea in these approaches is to find new field variables in which the anti-self-dual sector is linearized. These new field variables are related to the old ones by non-linear and non-local field transformations, expressed in the form of an infinite series. The possibility of such a transformation of the anti-self-dual sector into a free theory relies, in effect, on the complete integrability of the anti-self-dual Yang-Mills equations.  This leads to the triviality of the perturbative scattering theory, at least at tree level.  The MHV diagram formalism then provides the perturbation theory about this linearized anti-self-dual sector.

The Ward transform~\cite{Ward:1978wt} underlies the complete
integrability of the anti-self-dual Yang Mills equations, see for
example~\cite{Mason:1996mw}.  It linearizes the anti-self-dual
Yang-Mills equations by reformulating them as the condition that the
corresponding data on twistor space be holomorphic.  Our approach~\cite{Mason:2005zm, Boels:2006ir} builds on this foundation by writing
an action for the complete gauge theory (not merely its anti-self-dual interactions) on twistor space rather than on space-time. It uses the Euclidean space formulation of
the Ward construction~\cite{AHS1978eu}, in which twistor space, a
three dimensional complex manifold, is expressed as a 2-sphere bundle
over Euclidean space.  The Ward transform can then be understood in
terms of the possibility of choosing different gauges for the pullback
of the space-time Yang-Mills connection to twistor space.  As we show
explicitly in this paper, this allows us to pass between the standard
and the MHV descriptions of four-dimensional Yang-Mills perturbation
theory merely by making different gauge choices in the twistor
action. Furthermore, we show that our actions can be extended to
include fermionic or scalar matter coupled to the Yang-Mills field.
The same gauge choice then leads to a MHV diagram formalism for gauge
theories containing this matter in which the standard Yang-Mills MHV
diagrams are supplemented by extra towers of MHV diagrams
containing the matter interactions.  

We start with an introduction to the twistor formulation of Yang-Mills theory, and briefly describe the gauge choice that allows us to retrieve the standard space-time formulation.  We go on to discuss the gauge choice
that leads to the MHV diagram formalism and show how this comes about.
We provide extensions of the action to include scalar and fermionic
matter fields and show how the extra terms give rise to additional
towers of vertices in the MHV diagram formalism.  We give a brief
discussion of the connection between the standard space-time LSZ
formalism and that arising from the twistor action.  Questions still
remain, particularly concerning the origin of the MHV loop diagram.
We finish with a discussion of these and other matters.

\section{Twistor Yang-Mills}

The twistor space of four-dimensional Euclidean space $\bE$ may be
viewed as the total space of right Weyl spinors over $\bE$. A right
Weyl spinor $\pi_{\dot\alpha}$ in four dimensions has two components,
and we will be interested in the projectivized spin bundle where
$\pi_{\dot\alpha}$ is defined only up to the scaling
$\pi_{\dot\alpha}\sim \lambda\pi_{\dot\alpha}$ for any non-zero
complex $\lambda$. Thus projective twistor space $\bT'$ is a
$\bC\bP^1$ bundle over $\bE$, and may be described by coordinates
$(x^{\alpha,\dot\alpha},[\pi_{\dot\beta}])$ where $x\in\bE$ and
$[\pi_{\dot\beta}]$ is the point on the $\bC\bP^1$ with homogeneous
coordinates $\pi_{\dot\beta}$. The indices $\alpha,\dot\alpha=0,1$
denote the fundamental representation of each of the two $SL(2,\bC)$s
in the spin group $SL(2,\bC)\times SL(2,\bC)$ of the complexified
Lorentz group, and are raised and lowered using the
$SL(2,\bC)$-invariant tensors
$\epsilon_{\alpha\beta}=-\epsilon_{\beta\alpha}$ {\it etc.} We will
often employ the notation $\langle a\,b\rangle\equiv
\epsilon^{\dot\beta\dot\alpha}a_{\dot\alpha} b_{\dot\beta}$,
$[c\,d]\equiv\epsilon_{\beta\alpha}c^{\alpha}d^{\beta}$ and $\langle
a| M|c]\equiv a_{\dot\alpha}M^{\alpha\dot\alpha}c_{\alpha}$ to denote
these $SL(2,\bC)-invariant$ inner products. To preserve real Euclidean
space we restrict ourselves to the subgroup $SU(2)\times SU(2)\subset
SL(2,\bC)\times SL(2,\bC)$ which also leaves invariant the inner
products $\langle\pi\,\hat\pi\rangle$ and $[\omega\,\hat\omega]$,
where $(\hat\pi_{\dot 0},\ \hat\pi_{\dot
1})\equiv(\overline{\pi_{\dot1}},\ -\overline{\pi_{\dot0}})$ and
$(\hat\lambda^0,\ \hat\lambda^1)\equiv(\overline{\lambda^1},\
-\overline{\lambda^0})$.

Twistor space $\bT$ is the complex manifold $\bP^3$ with holomorphic
homogeneous coordinates
$(\omega^A,\pi_{A'})
=(x^{AA'}\pi_{A'},\pi_{A'})$, $(\omega^A,\pi_{A'})\sim
(\lambda\omega^A,\lambda\pi_{A'})$ and
$\bT'$ is the subset $\bP^3\backslash\bP^1$ on which $\pi_{A'}\neq 0$.  We will
work with coordinates $(x,\pi_{A'})$ on $\bT'$ for the most part, and
in these coordinates, the complex structure can be expressed in terms
of the following basis of (0,1)-forms and dual basis of
$(0,1)$-vectors adapted to the fibration over $\bE$
\begin{equation}
  \bar e^0 =
  \frac{\hat\pi_{\dot\alpha}\rd\hat\pi^{\dot\alpha}}
  {\langle\pi\,\hat\pi\rangle^2}\,  \qquad 
  \bar e^\alpha = \frac{\hat\pi_{\dot\alpha}\rd
    x^{\alpha\dot\alpha}}{\langle\pi\,\hat\pi\rangle} 
\qquad
\delbar_0 =
\langle\pi\,\hat\pi\rangle\pi_{\dot\alpha}\frac\del{\del\hat\pi_{\dot\alpha}}
\qquad \delbar_\alpha = \pi^{\dot\alpha}\frac\del{\del
x^{\alpha\dot\alpha}}
\label{basis}
\end{equation}
where the scale factors are chosen to ensure that the forms have only
holomorphic weight ({\it i.e.} they are independent of rescalings of
$\hat\pi$).   The $\delbar $ operator is defined on functions $f$ with
holomorphic weight to be $\delbar f= \bar e^0\delbar_0f +\bar e^\alpha
\delbar_\alpha f$, so that $\delbar f=0$ implies that $f$ is holomorphic
in $(\omega^\alpha,\pi_{\dot\alpha})$. 

To discuss space-time Yang-Mills theory on twistor space, we must
choose a vector bundle $E\to\bT'$ with vanishing first Chern class. Having $c_1(E)=0$ implies that $E$ is the pullback of some space-time bundle $\tilde E$ via the projection map, $E=\mu^*\tilde E$.  Our action will be a functional of fields $A\in\Omega^{0,1}_{\bT'}({\rm End} E)$ and
$B\in\Omega^{0,1}_{\bT'}(\cO(-4)\otimes{\rm End}E)$ where $A$ is
thought of as providing a $\delbar$-operator $\delbar_{gA}=\delbar+gA$
on $E$ and is defined up to appropriate gauge transformations $\delta
A=\delbar_{gA}\gamma$.  Here, $g$ is the Yang-Mills coupling
constant. We are not assuming that $\delbar_{gA}$ be integrable, so
that $\delbar_{gA}^2=gF$ is not assumed to be zero where $F=\delbar
A+g[A,A]$.

We then consider the action $S_{BF} + S_{B^2}$
where \cite{Mason:2005zm,Boels:2006ir}
\begin{equation}
S_{BF} =
\int_{\bT'}\!\!\!\Omega\wedge {\rm tr}(B\wedge F)\, \qquad
S_{B^2}:=-\frac12\int_{\bE\times \bP^1\times\bP^1}\hspace{-1cm} \rd^4x \,
\rD\pi_1\, \rD\pi_2\,
\langle\pi_1\pi_2\rangle^4{\rm tr}\left(B_1
K_{12} B_2 K_{21}\right) \label{action}
\end{equation}
The first term in this action is holomorphic $BF$ theory, with
$\Omega= \pi^{\dot\gamma}\rd\pi_{\dot\gamma}\wedge
\pi_{\dot\alpha}\pi_{\dot\beta}\rd x^{\alpha\dot\alpha}\wedge \rd
x^{\beta\dot\beta}\varepsilon_{\alpha\beta}$, the canonical top
holomorphic form of weight 4 on $\bT'$. This part of the action was
introduced by Witten \cite{Witten:2003nn} as part of a supersymmetric
Chern-Simons theory.  The second term is local on $\bE$ but is
non-local on the $\bP^1$ fibres. In it,
$K_{ij}=K_{ij}(x,\pi_i,\pi_j)$, $i,j=1,2$ are Green's functions for
the restriction $(\delbar_0+A_0)$ of the $\delbar_A$ operator to these
fibres described more explicitly in the next section, while
$\rD\pi_i=\langle\pi_i\,\rd\pi_i\rangle$ is the top holomorphic form
of weight 2 on the $i^{\rm th}$ fibre and $B_i=B(x,\pi_i)$ denotes $B$
evaluated on the $i$th factor.  This term is the lift to twistor space
of the $B^2$ term in the (space-time) reformulation of Yang-Mills
theory by Chalmers \& Siegel~\cite{Mason:2005zm,Chalmers:1996rq}.

The action is invariant under gauge
transformations
\begin{equation} 
\delta A=\delbar{\gamma}+ g[A,\gamma] \qquad\delta
B=g[B,\gamma]+\delbar\beta+g[A,\beta]\ , 
\label{gauge}
\end{equation}
where $\gamma$ and ${\beta}$ are smooth ${\rm End}\,E$-valued sections
of weight 0 and -4 respectively. 

In~\cite{Mason:2005zm,Boels:2006ir} we obtained the classical
equivalence of the twistor action with that on space-time by
(partially) fixing these gauge transformations by requiring $A_0=0$
and $B$ to be in a gauge in which it is harmonic over the $\bP^1$
fibres of $\bT'\to\bE$.  We refer to this as {\em space-time} gauge.
In this gauge $A$ can be expressed directly in terms of a space-time
gauge field $\cA_{\alpha\dot\alpha}$ and $B$ in terms of an auxilliary
space-time field $\cB_{\dot\alpha\dot\beta}$ and
the twistor action reduces to the
Chalmers and Siegel form of the usual Yang-Mills action on space-time 
\begin{equation}
\int_\bE \left(\cB_{\dot\alpha\dot\beta}\cF^{\dot\alpha\dot\beta} - \frac{1}{2}
\cB_{\dot\alpha\dot\beta}\cB^{\dot\alpha\dot\beta}\right)\rd^4 x 
\end{equation} with its usual space-time gauge symmetry.
Here, $\cF_{\dot\alpha\dot\beta}=
  \nabla_{(\dot\alpha}^\alpha\cA_{\dot\beta)\alpha} +
  \cA_{(\dot\alpha}^\alpha\cA_{\dot\beta)\alpha} $ is the self-dual
  part of the field strength of $\cA_{\alpha\dot\alpha}$ and
  $\cB_{\dot\alpha\dot\beta}$ is an auxiliary field that equals
  $\cF_{\dot\alpha\dot\beta}$ on shell.

The twistor action has a natural extension to supersymmetric gauge theory upto and including $\mathcal{N}=4$ and has an analogue for conformal
gravity~\cite{Mason:2005zm,Boels:2006ir}.

\section{Feynman Rules and MHV diagrams}
In this section, we will impose an axial gauge - first introduced by
Cachazo, Svr\v cek \& Witten~\cite{Cachazo:2004kj} - to obtain the
Feynman rules from (\ref{action}) and we will see that these directly
produce MHV diagrams. Note that the symmetry in~(\ref{gauge}), with
$\gamma$ and $\beta$ each depending on six real variables (the real coordinates of twistor space) is larger than the gauge symmetry of the
space-time Yang-Mills action; it is precisely by exploiting this
larger symmetry that we are able to interpolate between the standard
and MHV pictures of scattering theory.

Decomposing our fields into the basis (\ref{basis}) as $A=A_0\bar e^0 +
A_\alpha\bar e^\alpha$ {\it etc.}, we seek to impose the CSW gauge
condition $\eta^\alpha A_\alpha=0=\eta^\alpha B_\alpha$ for $\eta$
some arbitrary constant spinor. 
This is an axial gauge condition, so
the corresponding ghost terms will decouple. This gauge has the benefit 
that the $BAA$ vertex from the holomorphic $BF$ theory vanishes, being
the cube of three 1-forms each of which only has non-zero components
in only two of the three anti-holomorphic directions. Thus, the
only remaining interactions come from the non-local $B^2$ term. This
term depends on the field $A_0$ in a non-polynomial manner because of
the presence of the Green's functions $K=(\delbar_0+gA_0)^{-1}$ on the
$\bP^1$s. To
find the explicit form of the vertices, expand in powers of $A$ using
the standard formul\ae
\begin{equation}
\frac{\delta K_{12}}{\delta A}=\int_{\bP^1}\!\rD\pi_3\,
K_{13}gA(x,\pi_3)\!K_{32}
\quad\quad\hbox{and}\quad\quad
\left.K_{ij}\right|_{A=0}
=\frac{\bI}{2\pi\im}\frac{1}{\langle\pi_i\,\pi_j\rangle}\ \label{Kvariation}
\end{equation}
where $\bI$ is the identity matrix in the adjoint representation of
the gauge group.  Using this repeatedly gives the expansion
\begin{equation}
  K_{1p}=
  \sum_{n=1}^\infty \int_{(\bP^1)^{n-1}}\frac1{2\pi \im \langle \pi_n\,
    \pi_p\rangle}\prod_{r=2}^n \frac{g}{2\pi \im} \frac{A_r \wedge\rD\pi_r}{\langle \pi_{r-1}\, \pi_{r}\rangle} \label{Kexpansion}
\end{equation}
This can be substituted in to give the vertices
\begin{equation}
\sum_{n=2}^\infty\ \frac{g^{n-2}}{(2\pi\im)^n}\int_{\bE\times(\bP^1)^n}\hspace{-0.8cm}\rd^4x\
\left(\prod_{i=1}^{n}
\frac{\rD\pi_i}{\langle\pi_i\,\pi_{i+1}\rangle}\right)
\sum_{p=2}^n\langle\pi_{1}\,\pi_{p}\rangle^4 {\rm tr}\left(
B_1 A_2
\cdots A_{p-1}B_p A_{p+1}\cdots A_n
\right)\ . \label{vertices}
\end{equation}
This expression strongly resembles a sum of MHV amplitudes, except
that here we are dealing with {\it vertices} rather than amplitudes
and~(\ref{vertices}) is entirely off-shell.

In order to express the linear fields and propagators, it is helpful
to introduce certain $(0,1)$-form valued weighted $\delta$-functions
of spinor products
\begin{equation}
\bar \delta_{n-1,-n-1}\langle\lambda\,\pi\rangle
= \langle\hat\pi\,\rd\hat\pi\rangle\,
\frac{\langle\lambda\,\xi \rangle^n \langle\hat\lambda\,\xi \rangle} 
{\langle\pi\,\xi \rangle^n \langle\hat\pi\,\xi \rangle}\delta^2(\langle\lambda\,\pi\rangle)
\end{equation}
where for a complex variable $z=x+iy$,
$\delta^2(z)=\delta(x)\delta(y)$ and the scale factors have been
chosen so that $\bar\delta_{n-1,-n-1}\langle\lambda\pi\rangle$ has
holomorphic weight only in $\lambda$ and $\pi$ with weights $n-1$,
$-n-1$ respectively.  It is independent of the constant spinor
$\xi$ as $\lambda\propto\pi$ on the support of the delta function;
see \cite{Witten:2004cp} for a full discussion.  With these
definitions, $K_{ij}$ can be defined by
\begin{equation}
(\delbar_0+gA_0) K_{ij}=  \bI\bar\delta_{-1,-1}\langle \pi_i\,
\pi_j\rangle \, . \label{dbarK}
\end{equation}

The first term in~(\ref{vertices}) is quadratic in $B$ and involves
no $A$ fields. We are always free to treat such algebraic terms
either as vertices or as part of the kinetic energy of $B$. When
working in this axial gauge in twistor space, it turns out to be more
convenient to do the former, whereupon the only kinetic terms come
from the holomorphic $BF$ theory. In this axial gauge this is
\begin{equation}
\int_{\bT'}\!\Omega\wedge{\rm tr}(B\wedge F)
=\int_{\bT'}\!\Omega\wedge{\rm tr}(B\wedge\delbar A)
\label{freekinetic}
\end{equation}
so the propagator is the inverse of the $\delbar$ operator on $\bT'$.
Using coordinates $(x,\pi)$ on $\bT'$, the propagators depend on two
points, $(x_1,\pi_1)$, $(x_2,\pi_2)$, but, as usual, depend only on
the space-time variables through $x_1-x_2$. We can Fourier transform
the $x_1-x_2$ to obtain the momentum
space axial gauge propagator 
\begin{equation}
\langle A(p,\pi_1) B(p,\pi_2)\rangle 
=\frac{\bI}{p^2} 
\bar \delta_{-2,0} [\eta|p|\pi_1\rangle
\wedge \bar\delta_{2,-4} [\eta|p|\pi_2\rangle
+ \left( \frac {\bI}{\im}
\frac{\eta_\alpha\bar 
e^\alpha_1 \wedge \bar \delta_{2,-4} \langle\pi_1\pi_2\rangle 
}{[\eta|p|\pi_1\rangle} + 1\leftrightarrow 2 \right)
\label{propagators}
\end{equation}

The linearized field equations obeyed by external fields are $\delbar
A=0$ and $\delbar B=0$, while the linearized gauge transformations are
$\delta A=\delbar\gamma$ and $\delta B=\delbar\beta$. Together these
show that on-shell, free $A$ and $B$ fields are elements of the
Dolbeault cohomology groups $H^1_\delbar(\bT',\cO)$ and
$H^1_\delbar(\bT',\cO(-4))$ representing massless particles of
helicity $\mp1$ as is well-known from the Penrose transform.  Momentum
eigenstates obeying the axial gauge condition are
\cite{Cachazo:2004kj}
\begin{equation}
A(x,\pi)=T\e^{\im \tilde q_{\alpha}
x^{\alpha\dot\alpha}q_{\dot\alpha}} 
\bar\delta_{-2,0}\langle q\,\pi\rangle\qquad\qquad
B(x,\pi)=T\e^{\im\tilde q_{\alpha}
x^{\alpha\dot\alpha}q_{\dot\alpha}} 
\bar\delta_{2,-4}{\langle q\, \pi\rangle} \label{ffields}
\end{equation}
where $T$ is some arbitrary element of the Lie algebra of the gauge
group and $\tilde q_\alpha q_{\dot\alpha}$ is the on-shell momentum.  Only the
components $A_0$ and $B_0$ are non-vanishing and are simple multiples
of delta functions. Thus, inserting these external wavefunctions into
the vertices in (\ref{vertices}), one trivially performs the integrals
over each copy of $\bP^1$ by replacing $\pi_i^{\dot\alpha}$ by
$q_i^{\dot\alpha}$ for each external particle.  The integral over
$\bE$ then yields an overall momentum delta-function and, after colour
stripping the ${\rm tr}(T_1\ldots T_n)$ factors, one obtains the
standard form for an MHV amplitude, arising here from the Feynman
rules of the twistor action~(\ref{action}).

More general Feynman diagrams arise from combining the
vertices~(\ref{vertices}) with the propagators
from~(\ref{propagators}) and evaluating external fields
using~(\ref{ffields}).  This reproduces the MHV formalism for Yang-Mills
scattering amplitudes: the delta-functions in the propagator lead to
the prescription of the insertion of $[\eta|p$ as the spinor
corresponding to the off-shell momentum $p$. We note that the vertices
only couple to the components $A_0$ and $B_0$ of the fields so that
the second term in the propagator (\ref{propagators}) does not play a
role except to allow one to interchange the role of $B_0$ and $A_\alpha$.

We note that, since $\eta$ arises here as an ingredient in the gauge
condition, BRST invariance implies that the overall amplitudes are
independent of $\eta$.

\section{Coupling to Matter}

Matter fields of helicity $n/2$ correspond to $(0,1)$-forms $C_n$ of
homogeneity $n-2$ with values in ${\mathrm{End}}\,E$ for adjoint
matter, or $E$ or $E^*$ for `fundamental' matter.  These are subject to
a gauge freedom $C_n\rightarrow C_n+\delbar_{gA}\kappa_n$ where $\kappa_n$ is an arbitrary smooth function of weight $n-2$ with values in $E$, $E^*$ or ${\mathrm{End}}\,E$ as appropriate.  Thus an adjoint scalar field 
corresponds to a field $\phi$ on twistor space with values
in $\Omega^{0,1}({\mathrm{End}}\,E\otimes\cO(-2))$ and a fundamental
massless fermion corresponds to fields $\lambda$ in $\Omega^{0,1}(\cO
(-1))\otimes E)$ and $\tilde\lambda$ in $\Omega^{0,1}(\cO (-3))\otimes
E^*)$ subject to the gauge freedom
\begin{equation}\label{matter-gauge}
(\phi, \lambda,\tilde\lambda)\rightarrow 
(\phi+\delbar_{gA} \chi, \lambda+\delbar_{gA}\xi,\tilde\lambda+
  \delbar_{gA}\tilde\xi) 
\end{equation}
with $\chi, \xi,\tilde\xi$ of weights $-2,-1,-3$ respectively. The
local part of the matter action is
\begin{equation}
S_{\phi,\lambda,\tilde\lambda, \mathrm{Loc}}=
\int_{\bT'}\!\!\!
\Omega\wedge \left(\frac{1}{2}{\rm tr}\,\phi\wedge(\delbar\phi+g[A,\phi]) +
  \tilde\lambda\wedge(\delbar +gA)\lambda\right)   
\label{matter-action}
\end{equation}
It is clear that~(\ref{matter-action}) is invariant under the gauge transformations~(\ref{gauge}). However, in order for the sum $S_{BF}+ S_{\phi,\lambda,\tilde\lambda, \mathrm{Loc}}$ ({\it i.e.} the complete local action) to be invariant under~(\ref{matter-gauge}), we must modify the transformation rule of $B$ to
\begin{equation}\label{B-matter-gauge}
B\rightarrow B+\delbar\beta +g[A,\beta]+g[B,\gamma]-g[\chi,\phi]+ g\lambda
\tilde\xi-g\xi\tilde\lambda.
\end{equation}
This modified transformation law no longer leaves the $B^2$ term invariant, so we need to include new terms to compensate for
this additional gauge freedom. These terms may be found either by the Noether procedure. There is some freedom in the choice of these additional terms; we can fix this freedom by requiring that the action corresponds to matter fields which are minimally coupled on space-time. This requirement leads to the terms
\begin{eqnarray}
S_{B\, \phi^2 }&= &
g\int_{\bE}\hspace{-0.1cm} \rd^4x \, 
\int_{(\bP^1)^3}
\langle\pi_1\,\pi_2\rangle^2
\langle\pi_1\,\pi_3\rangle^2
{\rm tr} \left(B_1 K_{12}\phi_2K_{23}\phi_3 K_{31}\right)
\prod_{i=1}^3\rD\pi_i
\\
S_{\phi^4 }
&= &
g^2\int_{\bE}\hspace{-0.1cm} \rd^4x \, 
\frac12\int_{(\bP^1)^4}
\langle\pi_1\pi_2
\rangle^2\langle\pi_3\pi_4
\rangle^2\, {\rm tr} \left(
\prod_{i=1}^4
\, \phi_iK_{i,i+1}\rD\pi_i\right)
\\
S_{B\, \lambda,\tilde\lambda}&= &
g\int_{\bE}\hspace{-0.1cm} 
\rd^4x \, 
\int_{(\bP^1)^3}
\frac{\langle\pi_1\pi_2
\rangle\langle\pi_2\pi_3
\rangle^3}{\langle\pi_1\pi_3\rangle}\, 
\left(\tilde\lambda_1K_{12}B_2 K_{23}\lambda_3 
 \right) \prod_{i=1}^3\rD\pi_i 
\\
S_{\lambda^2,\tilde\lambda^2}&= &
g^2\int_{\bE}\hspace{-0.1cm} 
\rd^4x \, 
\int_{(\bP^1)^4}
\frac{\langle\pi_2\pi_4
\rangle\langle\pi_1\pi_3
\rangle^3
}{\langle\pi_4\pi_1
\rangle\langle\pi_2\pi_3
\rangle}
\left(\tilde\lambda_1K_{12}\lambda_2 \right)
\left(\tilde\lambda_3K_{34}\lambda_4\right)
\prod_{i=1}^4
\rD\pi_i
\\
S_{\lambda,\tilde\lambda\, \phi^2}&= &
g^2\int_{\bE}\hspace{-0.1cm} 
\rd^4x \, 
\int_{(\bP^1)^4}  \left[
\frac{
\langle\pi_4\pi_3\rangle
\langle\pi_1\pi_3
\rangle \langle\pi_1\pi_4
\rangle^2
}{\langle\pi_4\pi_1\rangle}
\left(\tilde\lambda_1K_{12}\lambda_2 \right)
{\rm tr} \left( \phi_3K_{34}\phi_4K_{43}\right) 
\right.\nonumber 
\\
&&\hspace{.7cm}\left.+
\frac{\langle\pi_1\pi_2\rangle^2 \langle\pi_1\pi_3\rangle
  \langle\pi_3\pi_4\rangle  + 2\leftrightarrow 3
}{2\langle\pi_4\pi_1\rangle}
\left(\tilde\lambda_1K_{12}\phi_2K_{23}\phi_3K_{34}\lambda_4 \right)
\right]
\prod_{i=1}^4\rD\pi_i 
\label{non-local-matter}
\end{eqnarray}
so that the total action is
\begin{equation}
S_{\rm Full}= S_{BF} + S_{\phi,\lambda,\tilde\lambda,
  \mathrm{Loc}} + S_{B^2} + S_{B\, \phi^2}+ S_{\phi^4}+ S_{B\, \lambda
  \, \tilde \lambda} +S_{\lambda^2\tilde\lambda^2 }
  +S_{\lambda,\tilde\lambda \phi^2}\ \ . 
 \label{full-action}
\end{equation}
To see that
these combinations are indeed invariant under~(\ref{matter-gauge}) and (\ref{B-matter-gauge}), note that $S_{B^2}$ will pick up a term which is an integral over $(\bP^1)^2$ of $g
\langle\pi_1\,\pi_2\rangle^4 {\rm tr }(B_1 K_{12}[\chi_2,\phi_2]K_{21})$
from the $\chi$ variation of $B$.  On the other hand, $S_{B\phi^2}$
will pick up a term from the variation of $\phi$ that is an integral
over $(\bP^1)^3$ of $g\langle\pi_1\,\pi_2\rangle^2
\langle\pi_1\,\pi_3\rangle^2 {\rm tr} \left(B_1 K_{12}\left(
(\delbar_{gA_2}\chi_2)K_{23}\phi_3 +
\phi_2K_{23}(\delbar_{gA_3}\chi_3) \right) K_{31}\right)$.  We can
integrate this last expression by parts, although care must be taken
because of the singularities in the integrand, and use (\ref{dbarK}) to
perform one of the $(\bP^1)^3$ integrals
to obtain $g\langle\pi_1\,\pi_2\rangle^4 {\rm tr }(B_1
K_{12}[\chi_2,\phi_2]K_{21})$ cancelling the corresponding term in the
variation of $S_{B^2}$. The $\chi$ part of the variation of $B$ in
$S_{B\phi^2}$ however leads to new terms that are in turn cancelled by
the corresponding variation of $S_{\phi^4}$ and so on for the $\xi$
and $\tilde\xi$ terms. 

All terms except $S_{B^2}$ vanish in space-time gauge, but are
necessary for the full gauge invariance of the action.  Thus the full
action corresponds on space-time to Yang-Mills with a minimally
coupled massless fermion $(\Lambda_\alpha,\tilde \Lambda_{\alpha'})$
in the (anti-)fundamental representation and a massless scalar field
$\Phi$ in the adjoint representation.  We note that we are free to
include  $\Phi^n$  interactions by including the additional gauge
invariant terms 
\begin{equation}
S_{\Phi^n}=
c_n\int_{\bE}\hspace{-0.1cm} \rd^4x \, 
\int_{(\bP^1)^4}
{\rm tr} \, \left(
\prod_{i=1}^n
\, \langle\pi_i\pi_{i+1}
\rangle
\phi_iK_{i,i+1}\rD\pi_i\right)
\label{phin}
\end{equation}
These terms are gauge invariant because the singularities in $K_{ij}$
in the integrand are cancelled by the factors of $\langle \pi_i\,
\pi_j\rangle$.  These correspond precisely to ${\rm tr }\Phi^n$ terms
in the space-time Yang-Mills theory by use of the standard integral
formula for scalar fields $\Phi(x)=\int_{\bP^1_x} H\phi
H^{-1}\,\rD\pi$ where $H$ is the gauge transformation to space-time
gauge.

As before, we can impose the CSW gauge condition $\eta^\alpha
A_\alpha=0=\eta^\alpha B_\alpha$, and similarly
$\eta^\alpha\phi_\alpha = 0 =\eta^\alpha\lambda_\alpha=
\eta^\alpha\tilde\lambda_\alpha$. In this gauge
\begin{equation}
\phi=T\e^{\tilde q_\alpha x^{\alpha\dot\alpha}q_{\dot\alpha}}
\bar\delta_{0,-2}\langle q\, \pi\rangle\, 
\quad \tilde\lambda=f\e^{\tilde q_\alpha x^{\alpha\dot\alpha}q_{\dot\alpha}}
\bar\delta_{-1,-1}\langle q\, \pi\rangle\, \quad
\lambda=f^*\e^{\tilde q_\alpha x^{\alpha\dot\alpha}q_{\dot\alpha} }
\bar\delta_{1,-3}\langle q\, \pi\rangle\, 
\end{equation}
where here $f$ is an element of the fundamental representation.  The
local parts of the action $S_{BF}+S_{\phi,\lambda\tilde\lambda {\rm
    Loc}}$ become quadratic and gives rise to propagators that have
the same form as~(\ref{propagators}), but with integers suitably
altered to reflect the different homogeneities. Just as for $S_B^2$,
each of $S_{B\phi^2}, S_{\phi^4}, S_{B\lambda 
\tilde\lambda}, S_{\lambda^2\tilde\lambda^2},
S_{\lambda,\tilde\lambda\phi^2}, S_{\Phi^n}$ can be seen to be
generating functions for MHV-type diagrams, now involving $\lambda,
\tilde \lambda$ and $\phi$, by expanding out the Green's functions $K_{ij}$ in powers of $A$ according to equation (\ref{Kexpansion}) and substituting into the
above formul\ae~as we did in (\ref{vertices}).  For example,
$S_{B\phi^2}$ gives rise to a sequence of MHV vertices each with one
positive helicity gluon corresponding to the $B$ field, two Higgs
fields $\phi$ and an arbitrary number of negative helicity gluons. The
other non-local terms in~(\ref{full-action}) behave similarly. The structure of these MHV-type vertices can be seen to be the same as was analysed in~\cite{Georgiou:2004by}.

\section{Connection to space-time calculation}

In space-time, Yang-Mills scattering amplitudes are calculated from
correlation functions using the LSZ formalism.  This instructs us to
calculate $\langle \cA_{\mu_1}(p_1) \ldots \cA_{\mu_n}(p_n)\rangle$,
isolate the residues of the single particle poles and contract with
the polarization vectors. The transform of the twistor-space field $A$
to a space-time gauge field $\cA$ is given by
\begin{equation}
\cA_{\alpha \dot\alpha}(x) =  \int k H \left( \delbar_\alpha +
A_{\alpha}  \right)H^{-1}  \frac{\hat{\pi}_{\dot\alpha}
}{\langle\pi\,\hat{\pi}\rangle} \label{spacetimeA}
\end{equation}
where $H$ solves $(\delbar_0+A_0)H=0$ and is the gauge transformation
to space-time gauge, and $k$ is a volume (K\"ahler) form on the
$\bP^1$. This follows from solving the constraint $F_{0\alpha}=0$
which arises by varying $B_\alpha$ in the action. Note that more than
one field insertion will lead to multiparticle poles and so can be
ignored at least at tree level. With this observation
the linearization of
(\ref{spacetimeA}) can be used so that, for the
application of the LSZ formalism,  operators
\begin{equation}
\cA_{\alpha\dot\alpha} =  \int k_1
\left(A_{\alpha}\frac{\hat{\pi_1}_{\dot\alpha}
}{\langle\pi_1\,\hat{\pi}_1\rangle} +\frac{\hat\pi_{1\,\dot\alpha}
  \delbar_{\alpha}}{\langle\pi_1\,\hat{\pi}_1\rangle}  \int k_2
\frac{\langle\pi_1\,\xi\rangle}{\langle\pi_2\,\xi\rangle}
\frac{A_0}{\langle\pi_1\, \pi_2\rangle}   \right) 
\label{eq:bungalowbill}
\end{equation}
should be inserted into the path integral. Here the second term is the
first term of the expansion of $H$ in $A_0$ and $\xi$ is an arbitrary
spinor reflecting the residual gauge freedom in
$\cA_{\alpha\dot\alpha}$. In the axial CSW gauge, the vertices that
contribute contain just fibre components of the $(0,1)$ forms.  The
only contractions to give $1/p^2$ poles are 
$\langle A_\alpha
A_0 \rangle$ and $\langle A_0 B_0\rangle$, the first of
these using one power of the $B^2$ vertex. The residues of these
contributions follow from~(\ref{eq:bungalowbill}) and are given
on-shell by
\begin{equation}
\int k_1 \left(A_{\alpha}\frac{\hat{\pi_1}_{\dot\alpha}
}{\langle\pi_1\,\hat{\pi}_1\rangle} \right) \rightarrow\ \eta_\alpha
[\eta\,\tilde q] q_{\dot\alpha}\qquad\qquad \int k_1
\left(\frac{\hat\pi_{1\,\dot\alpha}
\delbar_{\alpha}}{\langle\pi_1\,\hat{\pi}_1\rangle} \int k_2
\frac{\langle\pi_1\,\xi\rangle}{\langle\pi_2\,\xi\rangle}
\frac{A_0}{\langle\pi_1\, \pi_2\rangle} \right)\ \rightarrow\
\frac{\xi_{\dot\alpha} \tilde q_\alpha}{[\xi\, q] [\eta\,\tilde
q]^2}\ \ .
\end{equation}
These residues must be contracted with the polarization vectors which we recall are 
\begin{equation}
\varepsilon^-_{\alpha\dot\alpha} = \frac{ \tilde q_{\alpha} \kappa_{\dot\alpha}}{\langle\kappa\,q\rangle} \qquad\hbox{and}\qquad
\varepsilon^+_{\alpha\dot\alpha} = \frac{\tilde\kappa_{\alpha} q_{\dot\alpha}}{[\tilde\kappa\,\tilde q]}
\end{equation}
Hence it is clear that $A_\alpha$ and $A_0$ operator insertions correspond to insertions of different helicity states. In the above calculation $B_0$ could have been inserted instead of $A_\alpha$ since by the twistor transform on-shell $B_0$ is the field strength of the positive helicity gluon. Hence to insert the corresponding potential one can also use
\begin{equation}
\frac{p^{\alpha \dot\alpha} F_{\dot\alpha \dot\beta}}{p^2} = \frac{p^{\alpha \dot\alpha}}{p^2} \int k H B_0 H^{-1} \pi_{\dot\alpha} \pi_{\dot\beta}
\end{equation}
which can be verified to lead to the same prefactor as above. This is
in effect an expression of the field equation which relates $B_0$ and
$A_\alpha$. It is therefore seen that the usual calculation of
Yang-Mills amplitudes is equivalent to the prescription given earlier.

\section{Discussion}
The most important drawback of our (and indeed most other) approaches to the MHV formalism is that, while there are now a number of positive
results for loop amplitudes, it is still not clear that the extension
to loop level can be made systematic. This problem appears particularly acute in the
non-supersymmetric case where there are one-loop
diagrams that cannot be constructed from MHV vertices and propagators alone, see~\cite{Brandhuber:2006bf} for a full discussion.  Na\"ively it would seem
that such problems should not arise in our approach as the action
leads to a systematic perturbation theory including loops and the ghosts
decouple both in the space-time gauge and in the CSW axial gauge.  Nevertheless the missing one-loop diagrams provide us with a clear problem and it is still unclear how they might arise in our derivation of the MHV formalism. One possibility is that the missing diagrams arise from regularisation difficulties: it is well-known in
space-time that axial gauges require extra care for certain poles (see
{\it e.g.}~\cite{Leibbrandt:1987qv}), or the chiral nature of the action
may obstruct the implementation of an efficient regularisation scheme.
More likely however, in changing the path integral measure from
space-time to twistor space fields one encounters a determinant in the
spirit of~\cite{Feng:2006yy,Brandhuber:2006bf}. A potential source
for this determinant is the complex nature of our gauge transformation to the CSW gauge from the space-time gauge; the path integral is only invariant under real gauge transformations and such complex ones may require an extra determinant in the path integral measure.

Nonetheless, in our opinion this paper clearly shows that
the existence of the MHV diagram formalism can be understood in terms  of a linear and local gauge symmetry on twistor space.

\bigskip

\begin{acknowledgments}
  The authors would like to thank Freddy Cachazo and Wen Jiang for
  discussions. DS acknowledges the support of EPSRC (contract number
  GR/S07841/01) and a Mary Ewart Junior Research Fellowship. The work
  of LM and RB is supported by the European Community through the FP6
  Marie Curie RTN {\it ENIGMA} (contract number MRTN-CT-2004-5652).
\end{acknowledgments}


\begin{thebibliography}{99}

\bibitem{Witten:2003nn}
  E.~Witten,
  Commun.\ Math.\ Phys.\  {\bf 252}, 189 (2004)
  [arXiv:hep-th/0312171].

\bibitem{Cachazo:2004kj}
  F.~Cachazo, P.~Svrcek and E.~Witten,
  JHEP {\bf 0409}, 006 (2004)
  [arXiv:hep-th/0403047].

\bibitem{Bedford:2004nh}
  J.~Bedford, A.~Brandhuber, B.~Spence and G.~Travaglini,
  Nucl.\ Phys.\ B {\bf 712} (2005) 59
  [arXiv:hep-th/0412108].

\bibitem{Georgiou:2004by}
  G.~Georgiou, E.~N.~Glover and V.~Khoze,
  JHEP {\bf 0407} (2004) 048
  [arXiv:hep-th/0407027].

\bibitem{Birthwright:2005ak}
  T.~Birthwright, E.~N.~Glover, V.~Khoze and P.~Marquard,
  JHEP {\bf 0505} (2005) 013
  [arXiv:hep-ph/0503063].

\bibitem{Berger:2006vq}
  C.~Berger, Z.~Bern, L.~Dixon, D.~Forde and D.~Kosower,
  [arXiv:hep-ph/0607014].

\bibitem{Dixon:2004za}
  L.~Dixon, E.~N.~Glover and V.~Khoze,
  JHEP {\bf 0412} (2004) 015
  [arXiv:hep-th/0411092].

\bibitem{Badger:2005zh}
  S.~Badger, E.~N.~Glover, V.~Khoze and P.~Svrcek,
  JHEP {\bf 0507} (2005) 025
  [arXiv:hep-th/0504159].

\bibitem{Gorsky:2005sf}
  A.~Gorsky and A.~Rosly,
  JHEP {\bf 0601}, 101 (2006)
  [arXiv:hep-th/0510111].

\bibitem{Mansfield:2005yd}
  P.~Mansfield,
  JHEP {\bf 0603} (2006) 037
  [arXiv:hep-th/0511264].

\bibitem{Ettle:2006bw}
  J.~H.~Ettle and T.~R.~Morris,
  JHEP {\bf 0608} (2006) 003
  [arXiv:hep-th/0605121].

\bibitem{Feng:2006yy}
  H.~Feng and Y.~Huang,
  [arXiv:hep-th/0611164].

\bibitem{Brandhuber:2006bf}
  A.~Brandhuber, B.~Spence and G.~Travaglini,
  [arXiv:hep-th/0612007].

\bibitem{Ward:1978wt}
  R.~S.~Ward,
  Phys. Lett. B {\bf 61A} (1977) 81.

\bibitem{Mason:1996mw}
  L.~J.~Mason and N.~M.~J.~Woodhouse, Integrability, Self-duality and
  Twistor Theory, LMS monographs, (1996) OUP.

\bibitem{Mason:2005zm}
  L.~J.~Mason,
  JHEP {\bf 0510} (2005) 009
  [arXiv:hep-th/0507269].

\bibitem{Boels:2006ir}
  R.~Boels, L.~Mason and D.~Skinner,
  JHEP {\bf 0702} (2007) 014
  [arXiv:hep-th/0604040].

\bibitem{AHS1978eu}
M.~F.~Atiyah, N.~J.~Hitchin and I.~Singer, Proc. Roy. Soc. London,
{\bf A 362} (1978) 425-61.  


\bibitem{Chalmers:1996rq}
  G.~Chalmers and W.~Siegel,
  Phys.\ Rev.\ D {\bf 54} (1996) 7628
  [arXiv:hep-th/9606061].


\bibitem{Witten:2004cp}
  E.~Witten,
  Adv.\ Theor.\ Math.\ Phys.\  {\bf 8} (2004) 779
  [arXiv:hep-th/0403199].

\bibitem{Leibbrandt:1987qv}
  G.~Leibbrandt,
  Rev.\ Mod.\ Phys.\  {\bf 59} (1987) 1067.

\end{thebibliography}
\end{document}